1# Theoretical investigation of half-metallicity in Co/Ni substituted AlN

*Suneela Arif, Iftikhar Ahmad[*], B. Amin*

*Materials Modeling lab, Department of Physics, Garden Campus, Hazara University, Mansehra, Pakistan***Abstract**

Results of Co and Ni substituted AlN in the zinc blende phase are presented. For spin up states the hybridized N-2p and Co/Ni-3d states form the valance bands with a bandgap around the Fermi level for both materials, while in the case of the spin down states the hybridized states cross the Fermi level and hence show metallic nature. It is found that, $Al_{0.75}Co_{0.25}N$ and $Al_{0.75}Ni_{0.25}N$ are ferromagnetic materials with magnetic moments of 4 $\mu_B$ and 3 $\mu_B$ respectively. The integer magnetic moments and the full spin polarization at the Fermi level make these compounds half-metallic semiconductors. Furthermore it is also found that the interaction with the N-2p state splits the 5-fold degenerate Co/Ni-3d states into $t_{2g}$ and $e_g$ states. The $t_{2g}$ states are located at higher energies than the $e_g$ states caused by the tetrahedral symmetry of these compounds.

**Key words:** half metals, spintronics, spd-hybridization, AlN:Co, AlN:Ni

[*]     *Corresponding author's email address: ahma5532@gmail.com*

# 1   Introduction

Spintronics has received enormous attention [1-12] due to its interesting physics and potential applications in the high-tech spin based electronic devices. The effective use of spin along with charge of electrons in diluted magnetic semiconductors (DMS) makes them ideal candidates for spintronics applications. DMS can be synthesized by doping of transition metal (TM) atoms into a semiconductor crystal. These compounds are preferred over ferromagnetic materials due to their high spin injection efficiency and small mismatch in conductivity between them and the other components of the electronic devices [13].

Mn doped GaN is extensively investigated for spin based and optoelectronic devices. Many devices require a larger bandgap than the 3.07 eV of GaN [34]. AlN is a wide and direct bandgap compound with a bandgap of 5.4 eV [14]. It is extensively used as a host for optical devices [19]. Current applications of AlN include but are not limited to metal-insulator semiconductor hetro-structures, gas sensors, hetero-junction diodes, ultraviolet LED's and lasers [14-18]. AlN is also used as a substrate for GaAs and GaN and as a dielectric layer in optical storage devices. Due to its large bandgap it is expected that it can be also an ideal host for spintronic devices.

Sato et al. investigated $Al_{1-x}Co_xN$ for various Co concentrations of x; *0.06 - 0.25* [29]. Experimentally they found that the magnetization of the compound at low temperatures (*<50* K) increases with the increase in Co concentration from 17% to 25%. Li et al. reported ferromagnetism in Ni-doped polycrystalline AlN at 150 K, while in another study they found ferromagnetism at 300 K for a sample with a higher Ni concentration [20]. It has been experimentally reported that the Curie temperature of Co doped AlN is 300 K [35, 36]. Hebard et al. implanted Cr, Mn and Co in AlN crystals and described them ideal compounds for spintronics applications [30]. Recently Kaczkowski and Jezierski predicted ferromagnetism in Co/Ni (6.2%) doped AlN (wurtzite structure) using density functional theory [21].

In the present work, half-metallicity in $Al_{0.75}Co_{0.25}N$ and $Al_{0.75}Ni_{0.25}N$ in the zinc blende phase is investigated. The full potential linearized augmented plane wave (FP-LAPW) method with the generalized gradient approximation is used to calculate bandgaps and density of states of these compounds. The volume of the supercell with eight atoms of $Al_{1-x}(Co/Ni)_xN$ (Al is replaced by one atom of Co/Ni) is optimized with respect to the energy of the cell using Birch-Murangan's equation of state [37]. The equilibrium lattice constants are evaluated from these plots and are further used for the calculations of the electronic and magnetic properties of the compounds.

The substitution of one Al atom by a TM (Co/Ni) atom in the supercell, drastically changes the physical and chemical properties of the host crystal. The N-2s and TM-4s interact and form bonding and antibonding bands which are moved apart in energy. In addition there is hybridization between the TM-4s and TM-3d states. Moreover the ligand field generated by the N-2p states splits the 5-fold degenerate TM-3d states into doubly-degenerate $e_g$ and triply-degenerate $t_{2g}$ states, where the later are higher in energy.

## 2 Theory and calculations

The wave functions of a crystal can be effectively used to extract different physical properties of a compound. The solution of Schrödinger's equation for the many body problem of a crystal is extremely difficult, even with the fastest supercomputer. Hohenberg and Kohn [22] and Kohn and Sham [23] proposed the density functional theory (DFT) to simplify the problem, for which Kohn received the Noble prize in chemistry in 1998. They used the density functional instead of electron wave functions to solve the problem [22] and thus reduced 3N variables for N particles to three variables of the electronic density, i.e, ρ(r). The theory provides an exact solution to the problem. DFT is extremely reliable for the prediction and understanding of the structural, electronic, optical and magnetic properties of metals, semi-metals, semiconductors and insulators.

In the present work the Kohn-Sham equations are solved to calculate structural, electronic and magnetic properties of Co/Ni doped AlN crystal [23]. The Kohn-Sham equations in atomic units are:

$$[-\frac{1}{2}\nabla^2 + V_{ext}(\vec{r}) + V_C[\rho(\vec{r})] + V_{xc}[\rho(\vec{r})]\phi_i(\vec{r}) = \varepsilon_i \phi_i(\vec{r}) \qquad (1)$$

the first term on the left represents the kinetic energy operator, the second the external potential from the nuclei, the third is the Coulomb potential and the fourth one is the exchange correlation potential. The Kohn-Sham equations are solved iteratively till self consistency is achieved. Iteration cycles are needed because of interdependency between orbitals and potentials. In the Kohn-Sham scheme the electron density can be obtained by summing over all the occupied states [24].

The full potential linearized augmented plane-wave (FP-LAPW) method [25] with the generalized gradient approximation (GGA) [26] is used to solve Eq. 1. In the generalized gradient scheme, the exchange-correlation energy ($E_{xc}$) is a functional of the local electron spin densities $\rho(r)$ and their gradients:

$$E_{xc}^{GGA}(\rho_\uparrow,\rho_\downarrow) = \int \varepsilon_{xc}(\rho_\uparrow(\vec{r}),\rho_\downarrow(\vec{r}),\nabla\rho_\uparrow(\vec{r}),\nabla\rho_\downarrow(\vec{r}))\rho(\vec{r})d^3r \tag{2}$$

where $\rho_\uparrow$ and $\rho_\downarrow$ are densities for spin up and spin down electrons and $\varepsilon_{xc}$ is exchange correlation energy per particle. Details of the spin-polarized FP-LAPW method, formulas and the wien2k software used in the present calculations can be found in Ref. [27] and Ref. [28].

In the full potential scheme; radial functions times wave function, potential and charge density are expanded in two different basis sets. Within each atomic sphere the wave function is expanded in spherical harmonics while in the interstitial region it is expanded in a plane wave basis. In the same manner the potential is expanded as:

$$V(r) = \begin{cases} \sum_{lm} V_{lm}(r)Y_{lm}(\hat{r}) \ldots\ldots\ldots\ldots\ldots\ldots(a) \\ \sum_{K} V_{K} e^{iKr} \ldots\ldots\ldots\ldots\ldots\ldots\ldots(b) \end{cases} \tag{3}$$

Eq. 3(a) is for inside the sphere and 3(b) is for the interstitial region. The wave function is expanded in terms of spherical harmonics up to $l = 9$. Furthermore, inside the muffin tin sphere the potential is spherically symmetric and is constant elsewhere. The core electrons are treated fully relativistically and the valence electrons are treated semi-relativistically [27]. In order to ensure that no electron leakage is taking place semi-core states are included so that accurate results are achieved, 120 k points are used and $R_{MT} \times K_{max} = 8.00$ determines the plane wave basis functions.

## 3   Results and discussion

### 3.1   Electronic properties

In order to investigate the electronic properties of $Al_{0.75}Co_{0.25}N$ and $Al_{0.75}Ni_{0.25}N$, spin dependent density of states and spin dependent band structures of the compounds are calculated. Unlike rare-earth elements, where 4f electrons are partly shielded by the outer most 5s and 5p electrons to participate in the bonding, TM-3d electrons are exposed to the environment and can strongly be affected by the neighboring atoms. When a transition metal is substituted in a semiconductor host, it creates due to its high covalency an impurity level in the forbidden energy gap along with the permissible states. AlN is a wide bandgap semiconductor and the doping of Co/Ni affects its bandgap, as can be seen from Figs. 1, 2, 3 and 4.

The calculated band structures for the majority spin channels (spin up states) of $Al_{0.75}Co_{0.25}N$ and $Al_{0.75}Ni_{0.25}N$ are presented in Figs. 1(a) and 2(a). It is clear from the figures that there exist bandgaps of 2.6 eV and 3.0 eV respectively for these compounds. Furthermore the valence as well as conduction bands do not cross the Fermi levels and hence they behave like semiconductors. These gaps can also be clearly seen in the spin up state of the spin polarized total and partial density of states, presented in Figs. 3 and 4. These figures show that the valence bands are mostly occupied by Co/Ni -3d and N-2p states with a small contribution from the Al-s state. The valence bands of the compounds are very close to the Fermi level (but they do not cross the Fermi level). It can further be noted for the spin up states in Figs. 3 and 4, that the 2p states of N are located within the conduction bands with an energy gap of 2.6 eV for $Al_{0.75}Co_{0.25}N$ and 3.0 eV for $Al_{0.75}Ni_{0.25}N$ from the Fermi levels.

For the minority spin channels (spin down states) of $Al_{0.75}Co_{0.25}N$ and $Al_{0.75}Ni_{0.25}N$, the calculated spin polarized band structures are presented in Figs. 1(b) and 2(b). These figures show that the Fermi level crosses some of the valence band states and enters in the conduction band. This crossover reveals that these compounds are metallic in the spin down state. The origin of these bands can be clearly seen in the spin polarized total and partial density of states presented in Figs. 3 and 4. It is clear from these figures that the 3d states of Co and Ni crosses the Fermi level and enters the conduction bands.

In these compounds the N-2s and Co/Ni-4s states repel each other, shown in Fig. 5. Due to this repulsion among these states, the N-2s state is pushed towards the core and the TM-4s state towards the Fermi level. As the Co/Ni-4s and Co/Ni-4d states lie in the same energy range, hence they hybridize and form sd-hybridization. It can also be clearly seen in the figure that the hybridized Co/Ni-sd states are mixed with the N-2p states. The outcome of the overlapping of all these three states is spd-hybridization in these compounds. From the above discussion, based on the spin polarized band structures and density of states, $Al_{0.75}Co_{0.25}N$ and $Al_{0.75}Ni_{0.25}N$ are half metals.

The interesting feature of these compounds is the crystal field splitting. In the $Al_{0.75}(Co/Ni)_{0.25}N$ crystal, the cloud of p-orbitals of nitrogen splits the 5- fold degenerate Co/Ni-3d state into doubly degenerate, $e_g$ state, and triply degenerate, $t_{2g}$ state, shown in Figs. 3(b) and 4(b). For $Al_{0.75}Co_{0.25}N/(Al_{0.75}Ni_{0.25}N)$; Co-$e_g$/(Ni-$e_g$) is centered at -2.1 eV/(-2.06 eV) with a band width of 0.72 eV/(0.6 eV) and is located around 5.23 eV/(5.52eV) in the conduction band, while Co-$t_{2g}$/(Ni-$t_{2g}$) is centered at -0.30 eV/(-0.24) with a band width of 0.69 eV/(0.5 eV) and is located at 3.435 eV/(3.77 eV) from the conduction band edge. These calculations show that the wave function hybridizes very little for $e_g$ and strongly for $t_{2g}$ with the p state of N producing bonding and anti-bonding hybrids. The bonding hybrids lie within the valence band and the anti-bonding hybrids lie within the vicinity of the gap. The magnitude of the crystal field splitting for each compound can be calculated by: $\Delta E_{crystal} = E_{t2g} - E_{eg}$, where $E_{t2g}$ and $E_{eg}$ are the energies of the states $t_{2g}$ and $e_g$ respectively. Figs. 3(b) and 4(b) are used to estimate the crystal field splitting energies for both compounds. The calculated energies for $Al_{0.75}Co_{0.25}N$ and $Al_{0.75}Ni_{0.25}N$ are 1.87 eV and 1.75 eV respectively.

### 3.2 Magnetic Properties

The difference between the calculated ferromagnetic and anti-ferromagnetic energies, of $Al_{0.75}Co_{0.25}N$ and $Al_{0.75}Ni_{0.25}N$, show that both compounds are ferromagnetic in nature. The main source of magnetization in these materials is the unfilled 3d states. On further analysis of the d state it is the partly filled $t_{2g}$ state, which causes ferromagnetism in these compounds.

Ferromagnetism in III-V DMS is dominated by the double exchange interaction, which is a consequence of the large bandwidth and well localized wave function of the impurity states in the gap [38]. The substitution of TM (Co/Ni) atoms in the AlN crystal provides localized spins. The localized spin acts as an accepter center and generates a hole for the mediation of the ferromagnetic coupling between the TM spins, which causes double exchange interaction among the localized magnetic moments of these compounds [39, 40].

In the zinc blende lattices the d orbital of the transition metal splits, into triply degenerate $t_{2g}$ and doubly degenerate $e_g$ states, because of the tetrahedral crystal field. In tetrahedral crystal field the $e_g$ state lies lower in energy than the $t_{2g}$ state, in contrast to the octahedral field. So $t_{2g}$ hybridizes with the anion p-state and produces bonding and anti-bonding states, whereas the $e_g$ states do not participate in the hybridization process.

The data presented in Table 2 shows that the total magnetic moments of $Al_{0.75}Co_{0.25}N$ and $Al_{0.75}Ni_{0.25}N$ are the consequence of the contribution of Co/Ni, Al, N and interstitial sites. The main contribution to the net magnetic moment comes from the unfilled Co/Ni-3d states and considerable amounts from the cations and the interstitial sites. The table also shows that the local magnetic moment of Co reduces form its free space value of 3.00 $\mu_B$ to 2.58882 $\mu_B$ and for Ni it reduces from 2.00 $\mu_B$ to 1.64997 $\mu_B$. This decrease in the magnetization, on the atomic spheres of the TM atoms, is due to the induced magnetic moments on the non-magnetic sites. The positive values of the magnetic moments of Al, N and anti-sites (Table 2) show that their spins are aligned parallel to the TM spins. It is also evident from the table that the sums of all the individual magnetic moments for both compounds are integer values; 4.00 $\mu_B$ and 3.00 $\mu_B$ for $Al_{0.75}Co_{0.25}N$ and $Al_{0.75}Ni_{0.25}N$ respectively. One of the main properties of a compound to be a half metal is its integer magnetic moment [31-33]. The net integer magnetic moment also confirms that these compounds are half metallic ferromagnets.

4. **Conclusions**

It is concluded that the partly substitution of Al by Co or Ni abruptly changes structural, electronic and magnetic properties of AlN crystal. It is also concluded that $Al_{0.75}Co_{0.25}N$ and $Al_{0.75}Ni_{0.25}N$ are ferromagnetic substances. Furthermore, it is also concluded from the spin polarized density of states and spin polarized band structures of $Al_{0.75}Co_{0.25}N$ and $Al_{0.75}Ni_{0.25}N$ that these compounds are half metals.

## Acknowledgments

Prof. Dr. Nazma Ikram, (Ex) Director, Center of Excellence in Solid State Physics, University of the Punjab are acknowledged for her valuable suggestion and comments.

**Figure Captions**

**Fig. 1.** Spin polarized band structure of the supercell of $Al_{0.75}Co_{0.25}Ni$ (a) majority spin (b) minority spin

**Fig. 2.** Spin polarized band structure of the supercell of $Al_{0.75}Ni_{0.25}N$ (a) majority spin (b) minority spin

**Fig. 3.** Spin dependent total and partial density of states for the supercell of $Al_{0.75}Co_{0.25}N$

**Fig. 4.** Spin dependent total and partial density of states for the supercell of $Al_{0.75}Ni_{0.25}N$

**Fig. 5.** Repulsion of N-2s and Ni-4s states in the $Al_{0.75}Co_{0.25}N$ crystal

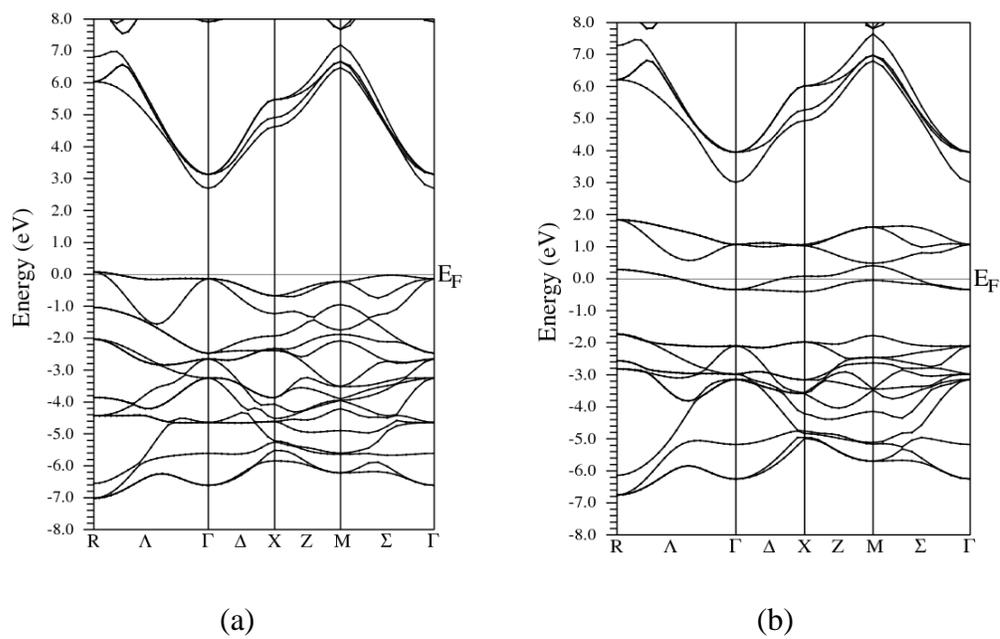

**Fig. 1.** Spin polarized band structure of the supercell of $Al_{0.75}Co_{0.25}Ni$ (a) majority spin (b) minority spin

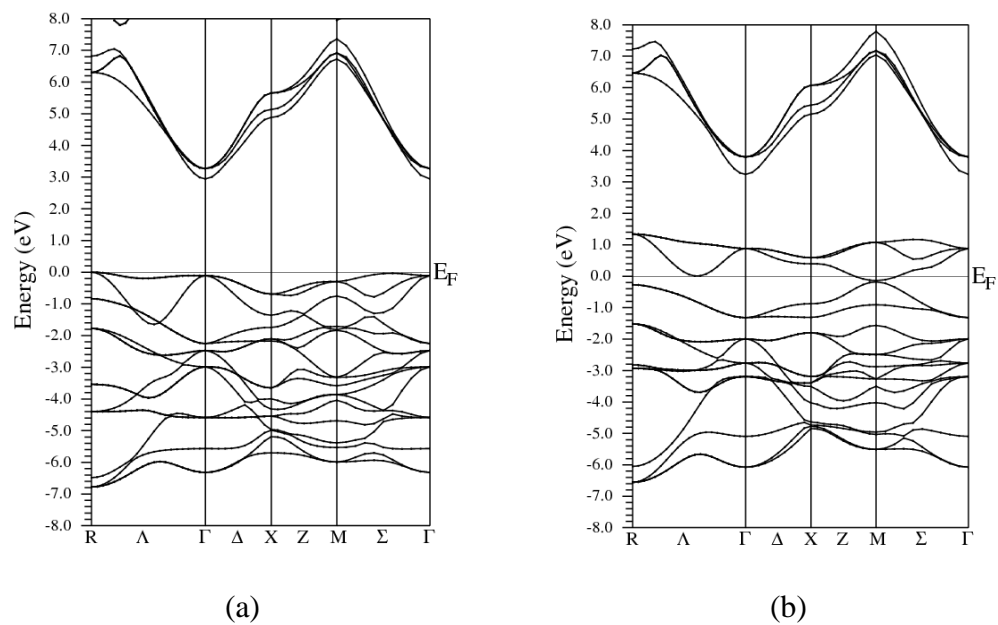

**Fig. 2.** Spin polarized band structure of the supercell of $Al_{0.75}Ni_{0.25}N$ (a) majority spin (b) minority spin

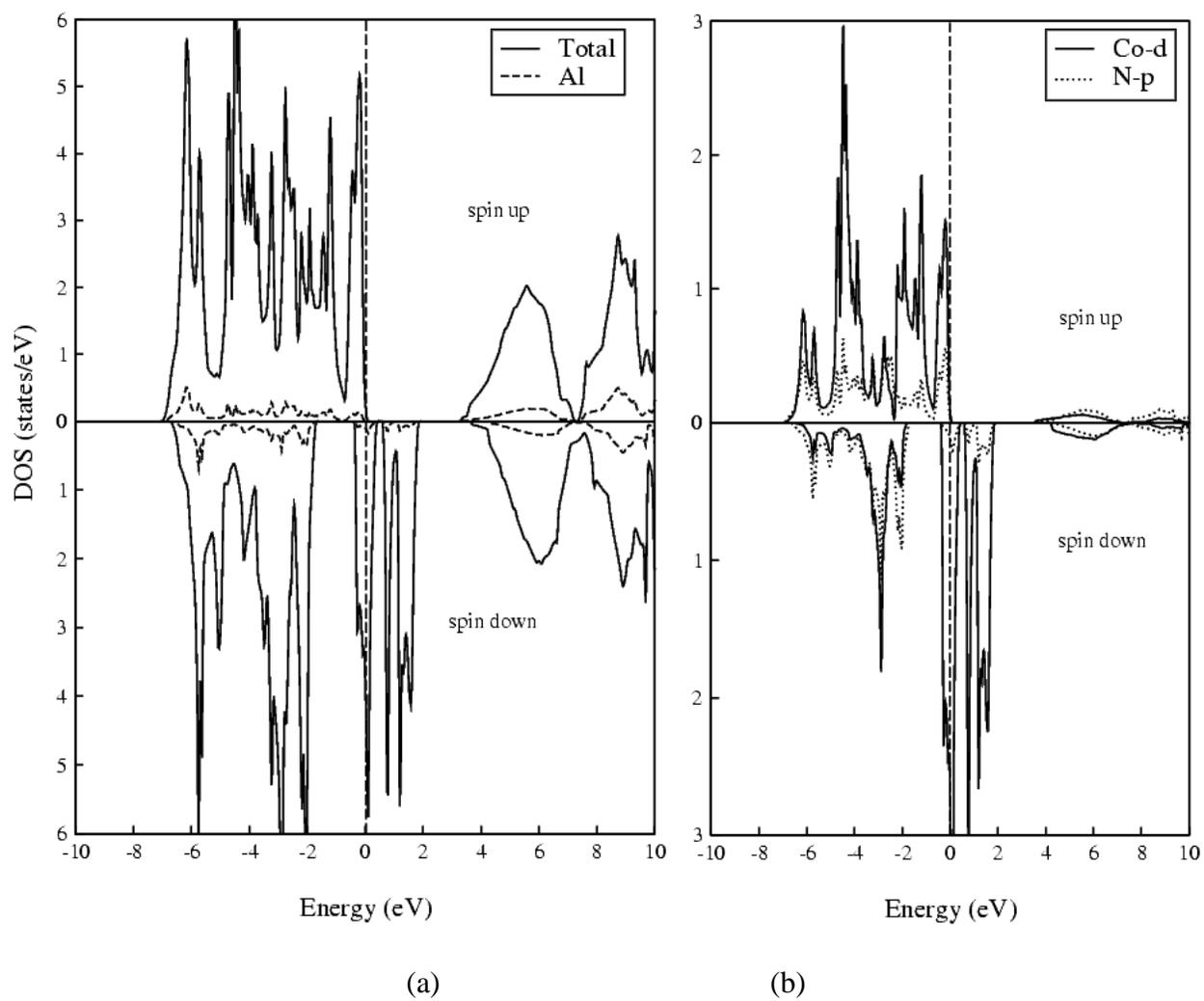

(a)            (b)

**Fig. 3.** Spin dependent total and partial density of states for the supercell of $Al_{0.75}Co_{0.25}N$

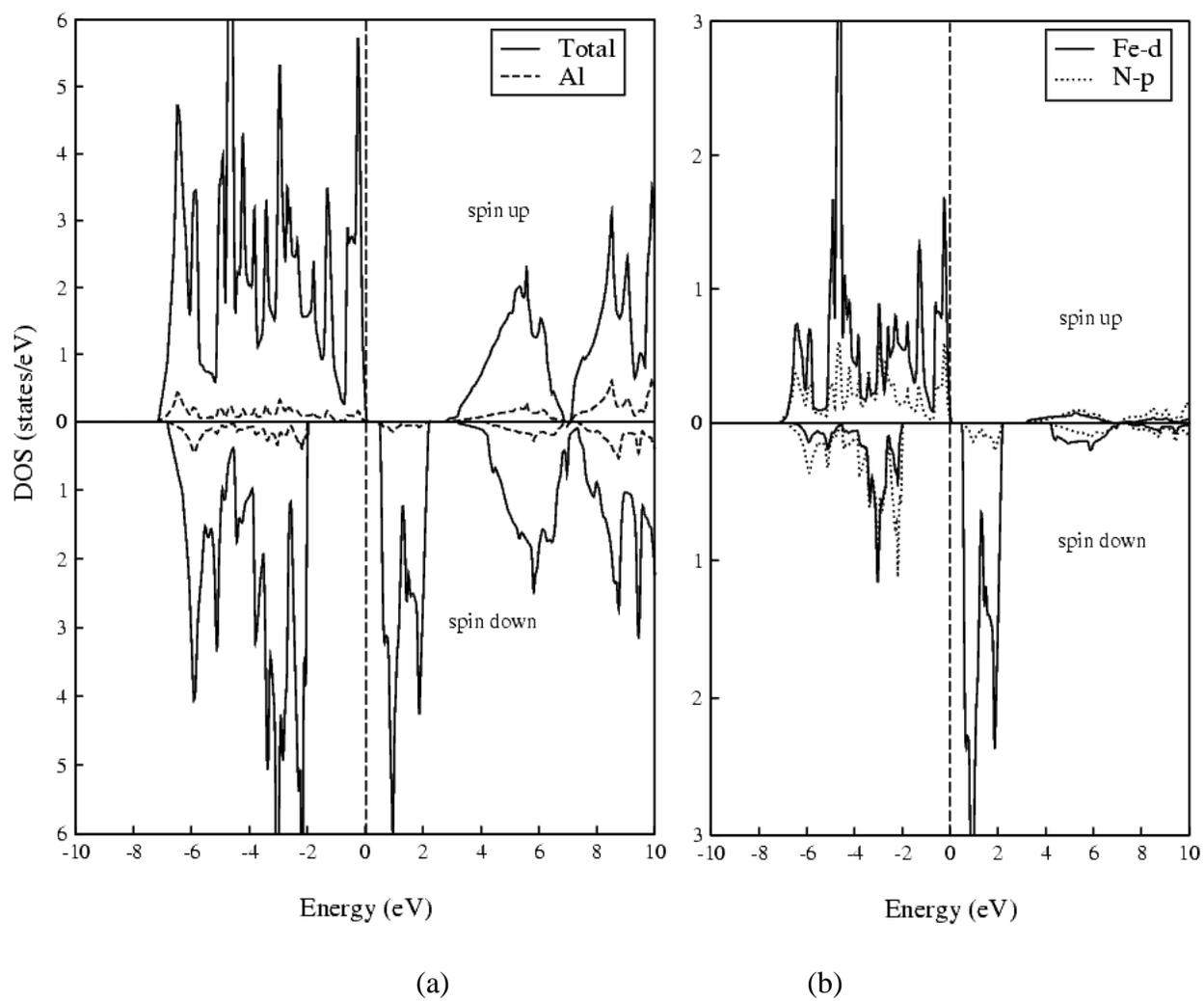

(a)          (b)

**Fig. 4.** Spin dependent total and partial density of states for the supercell of $Al_{0.75}Ni_{0.25}N$

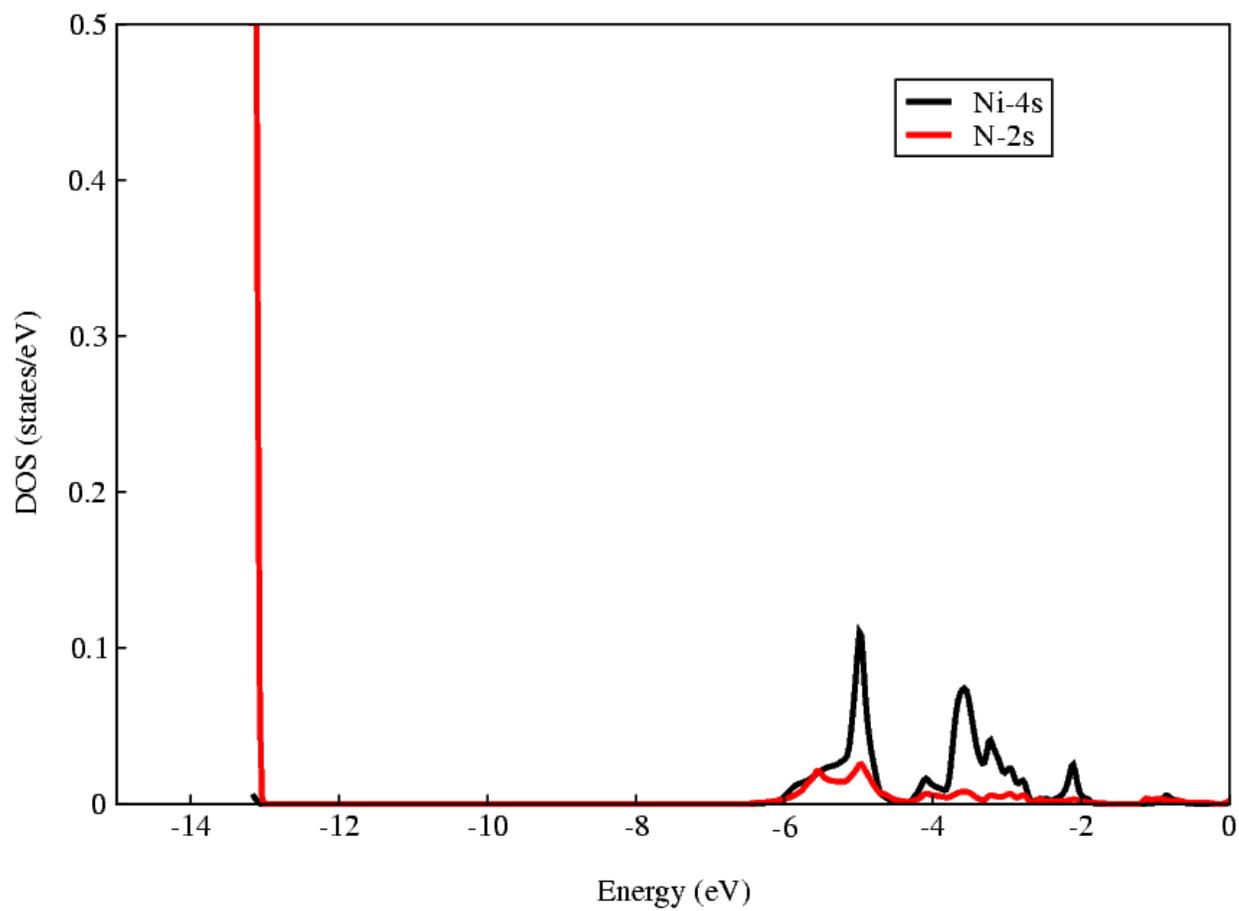

**Fig. 5.** Repulsion of N-2s and Ni-4s states in the Al$_{0.75}$Co$_{0.25}$N crystal

**Table 1**

Calculated lattice constants, bulk moduli and bandgap energies for $Al_{0.75}(Co/Ni)_{0.25}N$

| Compounds | Calculations | Lattice constant ($A^0$) | Bulk modulus (GPa) | Band-gaps (eV) |
|---|---|---|---|---|
| AlN | This work | 4.40 | 192.6064 | 5.4 |
|  | Other calc. |  |  | 5.4[a] |
| $Al_{0.75}Co_{0.25}N$ | This work | 4.4103 | 190.472 | 2.8 |
| $Al_{0.75}Ni_{0.25}N$ | This work | 4.407 | 191.3772 | 3.0 |

[a][14]

**Table 2**

Calculated total and local magnetic moments $\mu_B$ (in Bohr magneton) for $Al_{0.75}(Co/Ni)_{0.25}N$

| Site | $Al_{0.75}Co_{0.25}N$ | $Al_{0.75}Ni_{0.25}N$ |
|---|---|---|
| $M^{Tot}$ | 3.98748 | 3.00025 |
| $M^{Co}$ | 2.58882 | ---------- |
| $M^{Ni}$ | ---------- | 1.64997 |
| $M^{Al}$ | 0.02009 | 0.01375 |
| $M^{N}$ | 0.020681 | 0.21456 |
| $M^{int}$ | 0.51114 | 0.45078 |